# Localized Support for Injection Point Election in Hybrid Networks

Matthias R. Brust and Steffen Rothkugel
*Faculty of Science, Technology and Communication (FSTC)*
*University of Luxembourg*
*Luxembourg*
*{matthias.brust, steffen.rothkugel}@uni.lu*

## Abstract

*Ad-hoc networks, a promising trend in wireless technology, fail to work properly in a global setting. In most cases, self-organization and cost-free local communication cannot compensate the need for being connected, gathering urgent information just-in-time. Equipping mobile devices additionally with GSM or UMTS adapters in order to communicate with arbitrary remote devices or even a fixed network infrastructure provides an opportunity. Devices that operate as intermediate nodes between the ad-hoc network and a reliable backbone network are potential injection points. They allow disseminating received information within the local neighborhood. The effectiveness of different devices to serve as injection point differs substantially. For practical reasons the determination of injection points should be done locally, within the ad-hoc network partitions. We analyze different localized algorithms using at most 2-hop neighboring information. Results show that devices selected this way spread information more efficiently through the ad-hoc network. Our results can also be applied in order to support the election process for clusterheads in the field of clustering mechanisms.*

## 1. Introduction

Mobile devices such as cellular phones, notebooks, MP3-players, digital cameras, and alike become more commonplace in our daily life [1].

Multi-hop ad-hoc networks are composed of a set of these devices that communicate with each other over a wireless medium [2] using for instance Wi-Fi and Bluetooth.

Such networks can be established spontaneously whenever devices are in transmission range. In addition to *being mobile*, the desire of *being connected* with other devices or, in particular, the Internet, arises immediately [3].

The goal has been redefined to overcome the limitations of pure ad-hoc networks by augmenting them with instant Internet access. Technological advances like UMTS and GSM for linking to a backbone network drive considerable progress in this respect.

Devices offering both ad-hoc as well as backbone connectivity can operate as intermediate nodes between the different network types. We denote them using the term *injection points* [4]. The resulting network is of a hybrid nature.

Injection points serve two different purposes: a point where information dissemination starts [5] and where services are being placed [6].

In the first case, the injection point is of essential importance at the moment of receiving information and passing this information to the neighborhood.

The injection point might represent a bottleneck, depending on the amount of data passing through. Injection points become particularly attractive when offering a service. Information dissemination can be seen as such a service that is usable by devices in the injection point's surroundings.

However, it is necessary to pay particular attention on the efficiency of those injection points. Consequently, the dynamic and self-organizing augmentation of networks by additional links plays an important role. The costs of such augmentation should usually be made as small as possible.

Multiple different device properties have significant impact on the suitability of serving as injection point. Those include available power, technical equipment, load balancing issues, and for instance also the time a device is expected to remain available [7].

Aside from that, the effectiveness of injection points in terms of information spreading is crucial as well. The problem is to choose nodes that can disseminate information (in a region) more efficiently, i.e. with a minimized number of hops or messages.

For practical reasons the determination of injection points should be done locally, within the ad-hoc network partitions.

Thus, our contribution in this paper is to analyze several localized approaches to determine potential injection points. We apply this analysis to different network densities. We finally can offer guidelines on selecting the proper approach, depending on the network density at hand.

The approaches introduced use at most 2-hop topological information only, rather than relying on geographical positions.

The remainder of this paper is organized as follows. The underlying system model as well as the problem statement is illustrated in Section 2. In Section 3 localized approaches for the determination of injection points are introduced. We present an empirical study of these approaches in Section 4 and finally give a conclusion in Section 5.

## 2. Network model and problem statement

### 2.1. Hybrid network system model

A wireless network consists of a set of computers connected by wireless network links. We assume that technologies like Bluetooth and Wi-Fi can be employed to create ad hoc communication links within the transmission range at no charge. All nodes have the same communication range with bidirectional communication links.

Additional cellular network links such as GSM/UMTS might be employed by appropriately equipped devices to establish supplementary communication links to a network backbone. These links, however, will induce costs.

Our working assumption considers that the backbone network does not keep track of participating mobile devices. Thus, the request for information injection has to be initiated by a mobile device, the injection point candidate.

Note that we do not need to consider the details of the MAC and network layer for our investigations.

Furthermore we assume that each node knows its current one-hop neighbors and even the list of two-hop neighbors. Geographical positions of the nodes are not used at all.

### 2.2. Injection Point Candidate Problem

We define the injection point candidate problem as finding nodes that serve as more appropriate injection points according to some criteria than other nodes.

The objective is to classify nodes in order to indicate nodes that are more appropriate with high probability than the remaining nodes. These nodes are called injection point candidates.

## 3. Localized approaches

### 3.1. Bridge nodes

Bridge nodes connect two or more groups of nodes as illustrated in Fig. 1a. Thus, bridge nodes present a potential single-point-of-failure in terms of partitioning. These problematic situations appear when a node $v$ has a high number of neighbors and the neighbors are grouped. To illustrate this term suppose the graph $G=(N(v),E_N)$ is given where $N(v)$ is the set of neighboring nodes of $v$ and $E_N$ are the edges between the neighbors of $v$. If it is not possible to reach each node from any neighbor node as a starting point (ignoring node $v$), then we call the neighbors of $v$ grouped. In that case, the node v is called a bridge node, because it provides the only possible local path between these groups. Our assumption is that there is quantitive difference between efficiency of information spreading between bridge nodes and non-bridge nodes.

### 3.1. Weak nodes

*Weak nodes* are nodes that have less than three neighbors and all nodes with a clustering coefficient [8] less than a certain threshold $T_C$. Informally speaking, weak differentiate other nodes in some ways between sparse and dense regions of a network partition. The idea is that sparse regions may not be appropriate to inject information. The local clustering coefficient $CC$ of a node $v$ with $k_v$ neighbors is the number of links between the neighbors of $v$ divided by the number of all possible links which is $k_v(k_v-1)/2$.

### 3.2. (Obtrusive) Border nodes

A further approach supposes to place an injection point as central as possible in an ad-hoc network partition in order to reach remaining nodes "faster". Since it is difficult to detect the global center of topology only using local knowledge, we propose to exclude border nodes from the injection point candidates. *Border nodes* are located geographically at a border of a partition. The challenge in detecting border nodes lies in trying to estimate geographical relations while relying on local topological information only. The approach presented in [9] uses a reference node *r* for tackling this problem (cf. Fig. 1b).

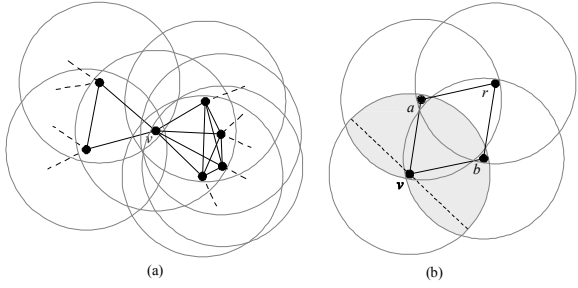

**Figure 1. Bridge node (a) and border node (b)**

Besides the reference node *r*, the method utilizes additionally two spanning nodes *a* and *b*. Assume a pair *a* and *b* exists, both neighbors of *v*, and both having a common neighbor that is not a neighbor of *v*. In this case, it will be checked if the neighborhood of *v* is completely covered by the neighborhood of *a* and *b*. If $\forall u \in N(v)$ with $u \in N(a \cup b)$ then *v* is marked as a border node. Ideally, the relationship between *r*, *a* and *b* is such that the resulting coverage (grey shaded area in Fig. 1b) covers around half of the coverage of *v*. Observe that the criterion for a border node is fulfilled if one set of nodes *a*, *b*, and *r* exists that covers the neighbors of *v* (obtrusive border nodes).

### 3.4. (Restrained) Border nodes

A more restrictive condition for border nodes is that all possible sets of nodes *a*, *b*, and *r* covers the neighbors of *v* (restrained border nodes). This restriction increases the number of injection point candidates.

### 3.5. Node degree

The node degree *d* of a node *v* is given by the number of its neighbors $|N(v)|$. Considering this, the node degree appears to be a very attractive heuristic since the node degree is 1-hop information and the minimum information available in an ad-hoc network. In contrast to, weak nodes, bridge nodes, and border nodes require 2-hop neighboring information.

## 4. Empiric Study

### 4.1. Experimental setting

**4.1.1. Topology.** As node deployment model we use the geometric random graph (GRG) model [10]. The GRG model describes a set of *n* nodes distributed according to a probability density function *pdf* in simulation area *R*. The links between nodes in *N* is based on a proximity relation as given by the transmission range.

For all experimental settings described in this paper we apply the uniform density function for node deployment in R.

**4.1.2. Network Density.** In order to detach from transmission range, simulation area, and the number of devices as simulation parameters, we define the parameter *network density*. The network density *d* is the sum of the total coverage area *Cov* of all nodes $n \in N$ divided by the simulation area *a*.

$$d = \frac{1}{a} \cdot \sum_{u \in N} Cov(RA(u))$$

Thus, the network density can be increased by either increasing the transmission range or the number of devices as well as decreasing the simulation area. We assume here a square area as simulation area as well as circular coverage areas.

Note there might be a more appropriate definition of network density for different settings as those assumed here.

We conducted all experiment for 30, 45, 60, 75, 90, 105, 120, 135, 150, 165, 180, 195, and 210 devices in a square area of 300×300 units while assuming a homogeneous transmission range of 50 units. These parameters correspondents to the network densities 2.62, 3.93, 5.24, 6.54, 7.85, 9.16, 10.47, 11.78, 13.09, 14.40, 15.71, 17.02, and 18.33.

**4.1.3. Experiments.** For the weak nodes approach, we conducted two experiments, one for $T_C = 0.35$ and one for $T_C = 0.4$ as threshold value.

Additionally, for the node degree approach two experiments, using the values 5 and 7 as thresholds, were conducted. Results shown in Fig. 2, Fig. 3, Fig. 4, and Fig. 5 are an average over 50 simulation runs.

## 4.2. Metrics

In order to find out if injection point candidates detected by the introduced approaches really tend to spread information more efficiently than any arbitrarily chosen node, we use a hop-count measurement. We understand a shorter path from one node to any other as a performance improvement of the information flow efficiency. The question of how to find this path is task of the routing layer that is not issue of this paper.

The all-pair shortest path $P$ of a network with a set of nodes $N$ is the average shortest path from any node to another.

$$P = \frac{1}{|N|(|N|-1)} \sum_{i \neq j} d(i,j) \quad \text{with } i, j \in N,$$

where $d(i,j)$ is the length of the shortest path between nodes $i$ and $j$. Assume the set of injection point candidates $I$, the injection point candidates-to-all shortest path is

$$P = \frac{1}{|I|(|I|-1)} \sum_{i \neq j} d(i,j) \quad \text{with } i \in I, j \in N.$$

Discharged candidates-to-all shortest path is calculated with the set $DC$ of discharged candidates respectively.

For this, the all-pair shortest path value is calculated and compared with the average one-to-all shortest path initiated by injection point candidates as well as with the average one-to-all shortest path initiated by discharged candidates, i.e. weak nodes, bridge nodes, border nodes, etc.

Observe that during this measurement we assume connected topologies, i.e. one partition, because the shortest path between two nodes is defined for connected nodes only. Therefore, we modified the topology generation slightly for this purpose. For this reason, we discharged partitioned topologies and considered the connected topologies. Due to the network density, in some cases topology discharge just occurred for densities 2.92 and 3.93 (border nodes and node degree).

Note also that there are different reasons why the network diameter represents an inappropriate metric here and the shortest path is a preferable metric. One reason is the average shortest path represents a characteristic for all nodes.

## 4.3. Results

We conducted experiments for each approach described in Section 3. For each experiment, we measured the all-pair shortest path, injection point candidates-to-all shortest path, and discharged candidates-to-all shortest path.

Bridge nodes are analyzed in Fig. 2. Region 1 shows a higher average path length for injection point candidates than for discharged candidates. The difference is approx. 10%. Thus, it seems that non-bridge nodes are very inappropriate injection point candidates, but that—in contrast—bridge nodes (discharged candidates) are appropriate for settings in region 1. The all-pair shortest path length is located between the values for injection point candidates and discharged candidates. This circumstance indicates that the number of injection point candidates and discharged candidates is similar.

Region 2 is confusing and it is not clear if there is a functional relationship between injection point candidates and discharged candidates. However, it is clear, that in the way the density is increasing the ratio between bridge nodes (i.e. in this case discharged candidates) and total number of nodes is decreasing. As result the values for average path length for injection point candidates and all nodes are getting very close.

Fig. 3 reveals results for weak nodes using $T_C = 0.35$ and $T_C = 0.4$ as threshold value. In both cases, region 1 shows that non-weak nodes are an appropriate choose as injection point candidate. Regions 2 comport similar to Region 2 in Fig. 2, indicating that the approach is not applicable for network densities used in the regions 2.

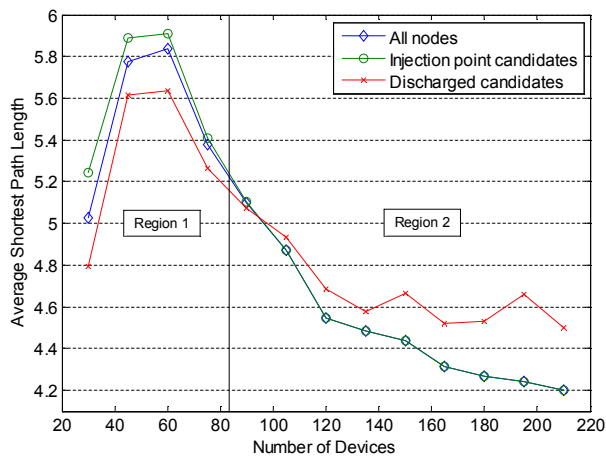

**Figure 2. Bridge nodes**

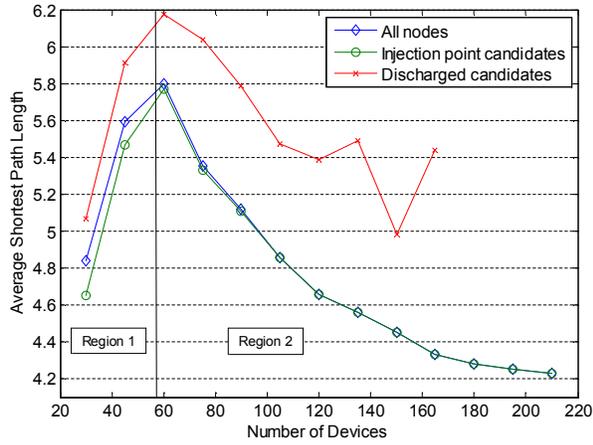
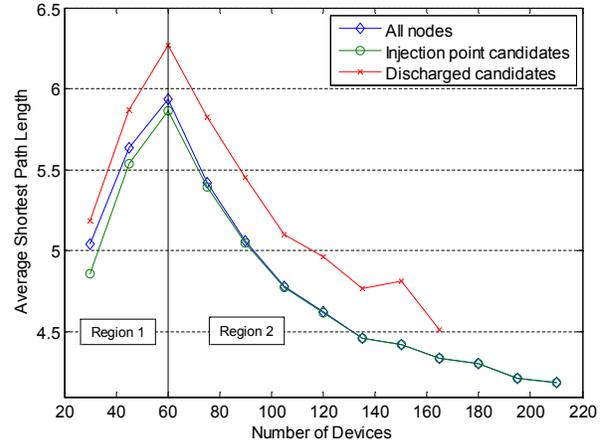

**Figure 3. Weak nodes with $T_C$ = 0.35 (left) and $T_C$ = 0.4 (right)**

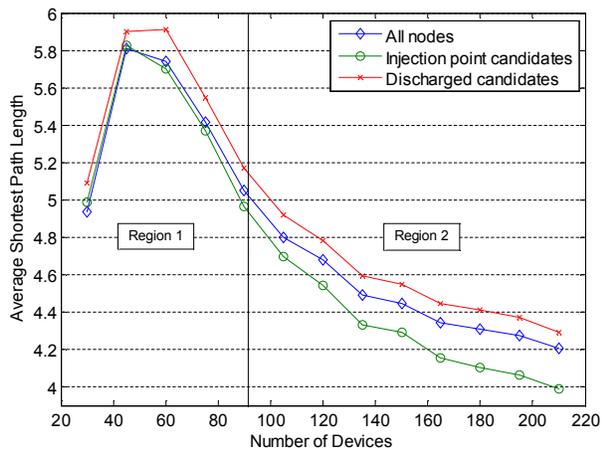
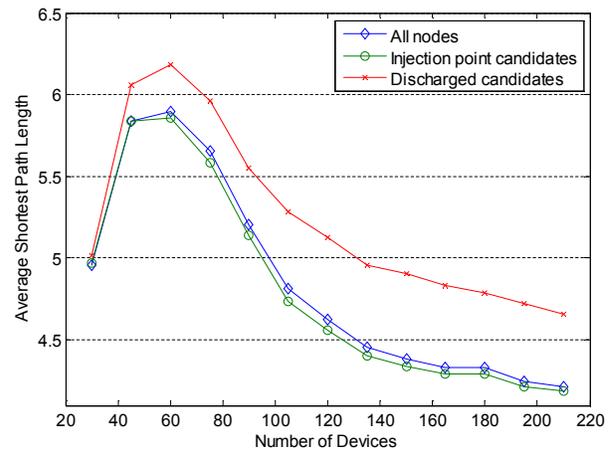

**Figure 4. Obtrusive border nodes (left) and restrained border nodes (right)**

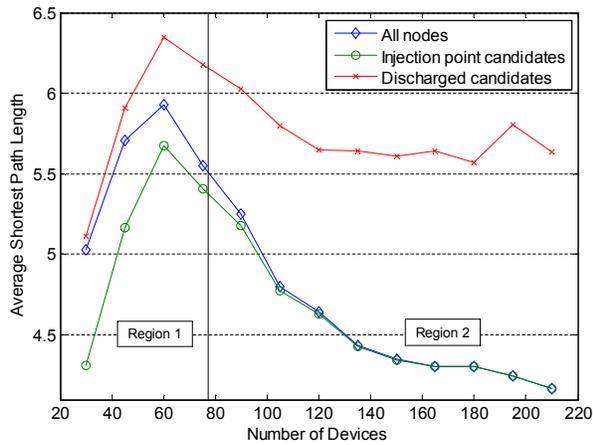
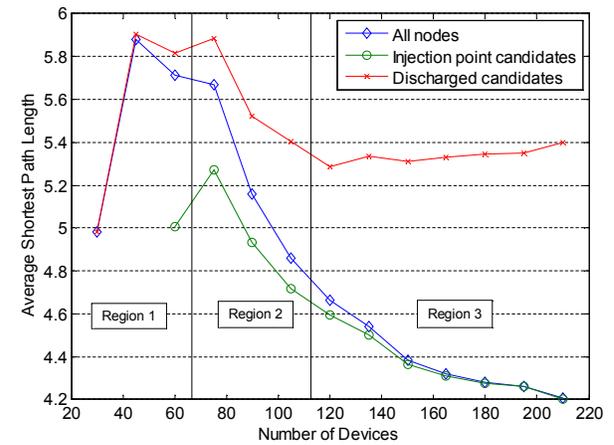

**Figure 5. Node degree for $k < 5$ (left) and $k < 7$ (right)**

Fig. 4 demonstrates that the border nodes approach performs acceptable for densities in region 2, gaining a hop improvement of approx. 10%. Region 1, however, does not reveal any benefits. In contrast to the obtrusive approach, the restrained approach seems to be fully inappropriate for all network densities.

Although the difference between path length of injection point candidates and discharged candidates is approx. 15%, the all-pair path length is very close to that of injection point candidates. Thus, the message complexity in order to detect restrained border nodes does not justify the small performance improvement gained by choosing an injection point candidate instead of choosing an arbitrary node as injection point. Furthermore, Fig. 4 implicates that the percentage of restrained border nodes is ignorable small.

The node degree approach with $k < 5$ and $k < 7$ has been analyzed in Fig. 5. In the case of $k < 5$ resulting injection point candidates perform approx. 15% better compared to the all-pair shortest path length.

This beneficial behavior drops down when increasing the network density (region 2). The reason is that the number of nodes with less than 5 neighbors decreases. For this, we conducted the same experiment using $k < 7$ as condition. Thus, region 2 reveals the beneficial network densities for injection point candidates with a similar improvement as in the case of $k < 5$.

## 5. Conclusion

Due to their characteristics, ad-hoc networks fail to work properly in a global setting, because these networks cannot compensate the need for being connected. This problem is tackled by introducing a backbone network link that is intended to be used on demand only. The mobile device that operates as intermediate node between ad-hoc network and backbone network is called injection point.

The effectiveness of different devices to serve as injection point differs substantially. For practical reasons the determination of injection points should be done locally, within the ad-hoc network partitions.

In this paper, we analyzed different localized algorithms using at most 2-hop neighboring information in order to support locally the injection point candidate election.

We applied this analysis to different network densities. The result can serve as guideline on selecting the proper approach, depending on the network density at hand.